\documentclass[twocolumn,amsmath,amssymb,prl,10pt,nofootinbib,superscriptaddress]{revtex4}
\usepackage{ graphicx, float,amsmath, amsmath, amssymb}
\usepackage[export]{adjustbox}

\def\be{\begin{equation}}
\def\ee{\end{equation}}
\def\bea{\begin{eqnarray}}
\def\eea{\end{eqnarray}}
\def\bse{\begin{subequations}}
\def\ese{\end{subequations}}

\usepackage[breaklinks, colorlinks, citecolor=blue]{hyperref}

\usepackage[labelsep=period]{caption}
\usepackage[normalem]{ulem}
\usepackage{soul}
\usepackage{minitoc}

\usepackage{bm}

\begin{document}
\title{Dark matter as Planck relics without too exotic hypotheses}

\author{Aur\'elien Barrau}%
\affiliation{%
Laboratoire de Physique Subatomique et de Cosmologie, Universit\'e Grenoble-Alpes, CNRS/IN2P3\\
53, avenue des Martyrs, 38026 Grenoble cedex, France
}

\author{Killian Martineau}%
\affiliation{%
Laboratoire de Physique Subatomique et de Cosmologie, Universit\'e Grenoble-Alpes, CNRS/IN2P3\\
53, avenue des Martyrs, 38026 Grenoble cedex, France
}



\author{Flora Moulin}%
\affiliation{%
Laboratoire de Physique Subatomique et de Cosmologie, Universit\'e Grenoble-Alpes, CNRS/IN2P3\\
53, avenue des Martyrs, 38026 Grenoble cedex, France
}

\author{Jean-Fr\'ed\'eric Ngono}%
\affiliation{%
Laboratoire de Physique Subatomique et de Cosmologie, Universit\'e Grenoble-Alpes, CNRS/IN2P3\\
53, avenue des Martyrs, 38026 Grenoble cedex, France
}


\date{\today}
\begin{abstract} 
The idea that dark matter could be made of stable relics of microscopic black holes is not new. In this article, we revisit this hypothesis, focusing on the creation of black holes by the scattering of trans-Planckian particles in the early Universe. The only new physics required in this approach is an unusually high-energy scale for inflation. We show that dark matter emerges naturally and we study the question of fine-tuning. We finally give some lines of thoughts for a possible detection.
\end{abstract}
\maketitle

\section{Introduction}

Dark matter is a very old problem. On the experimental side, it is being actively searched for, by direct detection (see, {\it e.g.}, \cite{Censier:2011wd,Gascon:2015caa,Mayet:2016zxu} for reviews), by indirect detection (see, {\it e.g.}, \cite{Cirelli:2012tf,Conrad:2014tla,Gaskins:2016cha} for reviews), and by accelerator production  (see, {\it e.g.}, \cite{Kahlhoefer:2017dnp,Felcini:2018osp} for reviews). Many ``little anomalies" are known, from the Fermi excess of GeV gamma rays \cite{TheFermi-LAT:2017vmf} to the PAMELA and AMS-02 overabundance of positrons \cite{Adriani:2008zr,Accardo:2014lma,Aguilar:2014mma}. All of them can however be quite simply accounted for by conventional astrophysical processes and at this stage no clear signal for nonbaryonic dark matter has been unambiguously recorded.\\

On the theoretical side, many hypotheses are being considered. They are actually too numerous to be exhaustively mentioned here (see, {\it e.g.}, \cite{Plehn:2017fdg} for an introductory review). From supersymmetry \cite{Bagnaschi:2015eha} to axions \cite{Klaer:2017ond}, most of them imply some amount of ``new physics". Recent developments even include an impressive list of highly speculative hypotheses.\\

Obviously, estimating the ``exoticity" of a model is quite subjective. In this brief article, we revisit the idea of dark matter made of Planck relics and we argue that this scenario might be much less exotic than most models. The only nonstandard hypothesis is a higher than usual reheating temperature.

\section{Trans-Planckian scattering}

Most studies considering primordial black holes (PBHs) are relying on production mechanisms that involve the collapse of overdense regions (see, {\it e.g.}, \cite{Carr:1975qj} for an early detailed calculation, \cite{Cline:1996mk,Jedamzik:1999am} for studies of phase transitions, and \cite{Carr:2009jm,Green:2014faa} for reviews). Those scenarios are however very unlikely as the density contrast required to form a PBH is close to 1 whereas the primordial power spectrum measured in the cosmological microwave background (CMB) has a much lower normalization. This bound could have been circumvented by a blue power spectrum as the scales involved in the formation of PBHs are much smaller than those probed by the CMB. The actual spectrum, however, happens to be red ($n_s\approx 0.965$) \cite{Aghanim:2018eyx}, making the production of primordial black holes by ``historical" mechanisms very difficult. Other scenarios like the collapse of cosmic strings were also considered \cite{Hawking:1987bn} but they are also disfavored -- if not ruled out -- by recent measurements. Interesting new ideas are, however, now being considered \cite{Garcia-Bellido:2019vlf,Carr:2019hud}.\\

Nevertheless, there exists a very different way to produce small black holes, namely through the scattering of trans-Planckian particles. As initially argued in \cite{Banks:1999gd}, when the impact parameter is smaller than the Schwarzschild radius (associated with the considered center-of-mass energy of a particle collision), the cross section for the scattering of trans-Planckian particles is dominated by an inelastic process leading to the formation of a single black hole. The key point is that the main features of high-energy scattering above the Planck energy can be studied from semiclassical considerations in general relativity (GR) and are therefore reliable. 

Basically, at impact parameters greater than the Schwarzschild radius, elastic and inelastic processes (gravitational
radiation, bremsstrahlung for charged particles, etc.) are described by solving the classical equations of the low-energy theory with initial conditions described by a pair of shock waves with appropriate quantum numbers. At smaller impact parameters, scattering is dominated by the “resonant” (in a sense different from the classical Breit-Wigner one) production of a  black hole with mass equal to the center-of-mass energy. The elastic cross section is suppressed by a Boltzmann factor and the incoming particles never get close enough together to perform a hard QCD scattering. In this limit the eikonal approximation for the initial state becomes valid and is described by a metric containing a pair of Aichelburg-Sexl shock waves with the associated impact parameter.

In \cite{Giddings:2001bu}, the study was refined and it was also concluded that the cross section for black hole production should be of the order of $\sigma(s)=F(s)\pi R_S^2(s)$ with $F(s)$ being a factor of order 1, $\sqrt{s}$ the center-of-mass energy, and $R_S$  the Schwarzschild radius. The details obviously depend on the considered quantum gravity theory but the main features are basically model independent.\\

Those ideas were applied to the possible production and observation of microscopic black holes at colliders (see, {\it e.g.}, \cite{Giddings:2001bu,Dimopoulos:2001hw,Barrau:2003tk,Paul:2005wb} for early works) in theories with a low Planck scale -- typically in the TeV range (usually associated with the existence of large extra dimensions \cite{ArkaniHamed:1998rs} or with many new particle species \cite{Dvali:2007hz}). A nice review including astrophysical effects, like those mentioned in \cite{Barrau:2005zb}, can be found in \cite{Kanti:2004nr}. In this article, we do not rely on the existence of extra dimensions and we do not assume a low Planck scale. 

\section{Stable relics}

The Hawking temperature $T_H=1/(8\pi M)$ \cite{Hawking:1974sw} is vanishingly small for astrophysical black holes but becomes significant for very small black holes. The mass loss rate during the evaporation is proportional to $M^{-2}$ and the process is therefore highly explosive. In itself, the evaporation mechanism is well understood from many different perspectives and is very consensual (see, {\it e.g.}, \cite{Lambert:2013uaa} for a simple introduction). Although it has not been observationally confirmed, there are indications that it might have been revealed in analog systems \cite{Steinhauer:2015saa}. \\

The status of the end point of the evaporation process is less clear. Obviously, the semiclassical treatment breaks down in the last stages and the divergence of the temperature together with the appearance of a naked singularity is nonphysical. Many different arguments have been pushed forward in favor of the existence of stable Planck relics at the end of the evaporation process (see \cite{Barrow:1992hq,Zeldovich:1983cr,Aharonov:1987tp,Banks:1992ba,Banks:1992is,Bowick:1988xh,Coleman:1991jf,Lee:1991qs,Gibbons:1987ps,Torii:1993vm,Callan:1988hs,Myers:1988ze,Whitt:1988ax,Alexeyev:2002tg} to mention only a few historical references, among many others). There are excellent arguments from quantum gravity, string gravity or modified gravity theories in favor or remnants. Those are however obviously based on ``new physics". One of the best arguments for Planck relics using only known physics was given by Giddings in  \cite{Giddings:1992hh}. Locality, causality and energy conservation considered within the information paradox framework (see, {\it e.g.}, the first sections of \cite{Mathur:2012np} for a precise description) do suggest that the time scale for the final decay of BHs is larger than the age of the Universe.\\

Although no clear consensus  exists on the status of BHs at the end of the evaporation process, it is fair to suggest that the existence of relics is somehow simpler from the viewpoint of usual physics. A recent review on the pros and cons of stable remnants can be found in \cite{Chen:2014jwq}. It is concluded that if relics contain a large interior geometry -- which is supported by \cite{Christodoulou:2014yia,Christodoulou:2016tuu} --, they help solve the information loss paradox and the firewall controversy.

\section{Reheating scale}

The idea that dark matter could be made of Planck relics was first suggested in \cite{MacGibbon:1987my}. This seminal work was, however, focused on PBHs formed by the collapse of overdense regions (or similar mechanisms), which is now  believed to be extremely unlikely as previously pointed out. We focus here on the possibility that PBHs are formed by the collision of trans-Plankian particles in the early Universe. This has already been considered in \cite{Saini:2017tsz} and in \cite{Conley:2006jg,Nakama:2018lwy} (see also references therein) for the case with extra dimensions.\\

In this work, we so not assume a lower than usual Planck scale due to extra dimensions. We quite simply consider the standard cosmological scenario in a (3+1)-dimensional spacetime and just take into account the ``tail" of trans-Planckian particles at the reheating time. The key point lies in the fact that the potentially produced relics behave nonrelativistically and are therefore be much less diluted (their energy density scales as $a^{-3}$) than the surrounding radiation (whose energy density scales as  $a^{-4}$). Hence, it is possible to reach a density of relics (normalized to the critical density) close to 1, $\Omega_{rel}\equiv\rho_{rel}/\rho_{cr}\approx 1$, with only a tiny fraction of relics at the formation time. The relative ``amplification" of the relics density compared to the radiation density between the reheating and the equilibrium times is given by $T_{RH}/T_{eq}\approx 3\times 10^{27} T_{RH}$ when $T_{RH}$ is given in Planck units. To fix ideas, for a reheating temperature at the GUT scale, a relics fraction of only $10^{-24}$ at the formation time would be enough to nearly close the Universe  at the equilibrium time. \\

For a thermal distribution of particles at temperature $T$, the number of particles above $E_{th}>T$ is exponentially suppressed. This is why, even with the amplification factor given above, the scenario presented here requires a reheating temperature not much below the Planck scale. This constitutes, in our view, the only ``nonstandard" input of this model. The Planck experiment final results lead to an upper limit on the tensor-to-scalar ratio of primordial perturbations $r<0.1$ \cite{Akrami:2018odb} which is even tightened to $r<0.064$ by combining the data with the BICEP2/Keck Array BK14 measurements. This is usually interpreted as an upper limit on the energy scale of inflation around the GUT scale (the higher the energy scale, the larger the normalization of tensor modes), which is too low for the process considered here. There are, however, at least two ways to circumvent this bound (we assume for simplicity a sudden reheating). 

The first one consists in noticing that the upper limit on the energy scale of inflation holds firmly only for rudimentary models. In $k-$inflation \cite{ArmendarizPicon:1999rj}, the relation basically becomes $r=-8C_Sn_t$ (instead of $r=-8n_t$), where $n_t$ is the tensor index and $C_S<1$ is the speed of sound for perturbations. This relaxes the bound. In two-field inflation \cite{Wands:2007bd}, the upper limit is also relaxed to $r=-8n_tsin^2(\theta)$, where $\theta$ accounts for the possible evolution of adiabatic scalar modes on super-Hubble scales. In multifield inflation the relation between $r$ and $n_T$ even becomes an inequality.

A second and probably more provocative argument would be the following. Whereas temperature anisotropies originate from usual quantum physics, namely from the quantum fluctuations of the inflaton field, the tensor perturbations leading to B modes in the CMB should come from the quantum fluctuations of the polarization modes of the graviton. In a sense (and although some counterexamples have been constructed but for ``artificial" models), B-modes would be a signature of perturbative quantum gravity (dimensional arguments are given in \cite{Krauss:2014sua}). Quantum gravity is a fascinating area of research but it has still no connection with experiments and assuming gravity not to be quantized is also legitimate, especially when considering how difficult and paradoxical is the quantization of the gravitational field \cite{Oriti:2009zz}. It is therefore meaningful to consider the possibility that {\it no} B mode is produced, even with a very high-energy scale for inflation, just because gravity might not be quantum in nature (this would also raise many consistency questions but is obviously worth being considered, as advocated in \cite{Jacobson:1995ab,Eling:2006aw}). In such a case, the usual upper bound could also be ignored.\\

Obviously, the normalization of the scalar spectrum would also be in tension with such a high scale (violating the slow-roll conditions in the most simple cases).
We do {\it not} mean that a higher than usual energy scale for inflation is unavoidable or even favored. We simply state that this is not ruled out by the tensor-to-scalar ratio and might be, in our opinion, less ``exotic" than most assumptions required for usual DM candidates. 

\section{Dark matter abundance}

The threshold energy $E_{t}$ to produce a BH in a head on collision of particles is expected to be of the order of the Planck energy but, depending on the details of the considered model, might be slightly different and we keep it as a free parameter. To estimate the number density of particles above $E_{t}$, one simply needs to integrate the thermal distribution, which leads to
$$n_{part}\approx T_{RH}e^{-{E_{t}/T_{RH}}},$$
where we use Planck units (as everywhere in this work except otherwise specified). Obviously, if the reheating temperature is too small when compared to the threshold energy of BH production, the number of PBHs will be exponentially suppressed and the process will be inefficient. The cross section, in principle, depends on the energy of the collision but, as a fist step, can be assumed to be a constant $\sigma_{BH}$ above the threshold. The collision rate is therefore given by $\Gamma=n_{part}\sigma_{BH} v\approx n_{part}\sigma_{BH}$. The energy density of radiation is
$$\rho_R=\frac{\pi^2}{30}g_*T^4_{RH},$$
with
$g_*$ being the total number of effectively massless degrees of freedom, that is species with masses $m_i\ll T_{RH}$. The Hubble parameter is
$$H=1.66g_*^{1/2}T_{RH}^2.$$ If relics are assumed to have a mass $m_{rel}$ (necessarily  lower than $E_{th}$), the energy density of relics is given by
$$\rho_{rel}\approx\frac{n_{part}m_{rel}\Gamma}{H}\approx\frac{e^{-\frac{2E_t}{T_{RH}}}\sigma_{BH}m_{rel}}{1.66g_*^{1/2}}.$$
The relative density of relics at the formation time is 
$$\Omega_{rel}^f=\frac{30\sigma_{BH}m_{rel}}{1.66\pi^2g_*^{3/2}}\cdot\frac{e^{-\frac{2E_t}{T_{RH}}}}{T_{RH}^4},$$
leading, in agreement with \cite{Nakama:2018lwy}, to a relative density at the equilibrium time of
$$\Omega_{rel}^{eq}=\frac{30\sigma_{BH}m_{rel}}{1.66\pi^2g_*^{3/2}}\cdot\frac{e^{-\frac{2E_t}{T_{RH}}}}{T_{eq}T_{RH}^3}.$$

\begin{figure} 
\centering
    \includegraphics[width=1.1\linewidth]{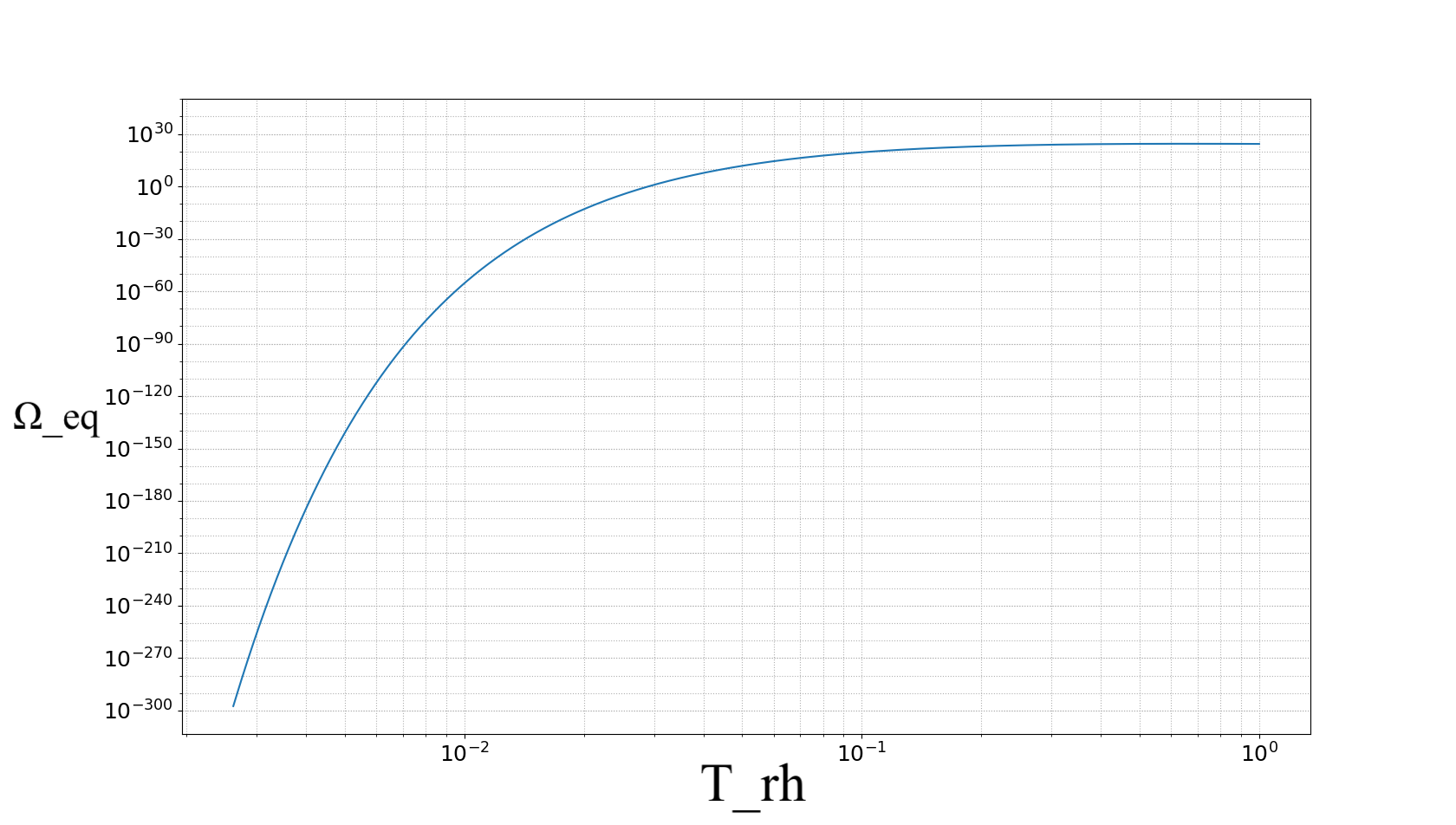}  
 \caption{Fraction of relics at the equilibrium time as a function of the reheating temperature (in Planck units).} 
      \label{Fig1}
\end{figure}

\begin{figure} 
\centering
    \includegraphics[width=1.1\linewidth]{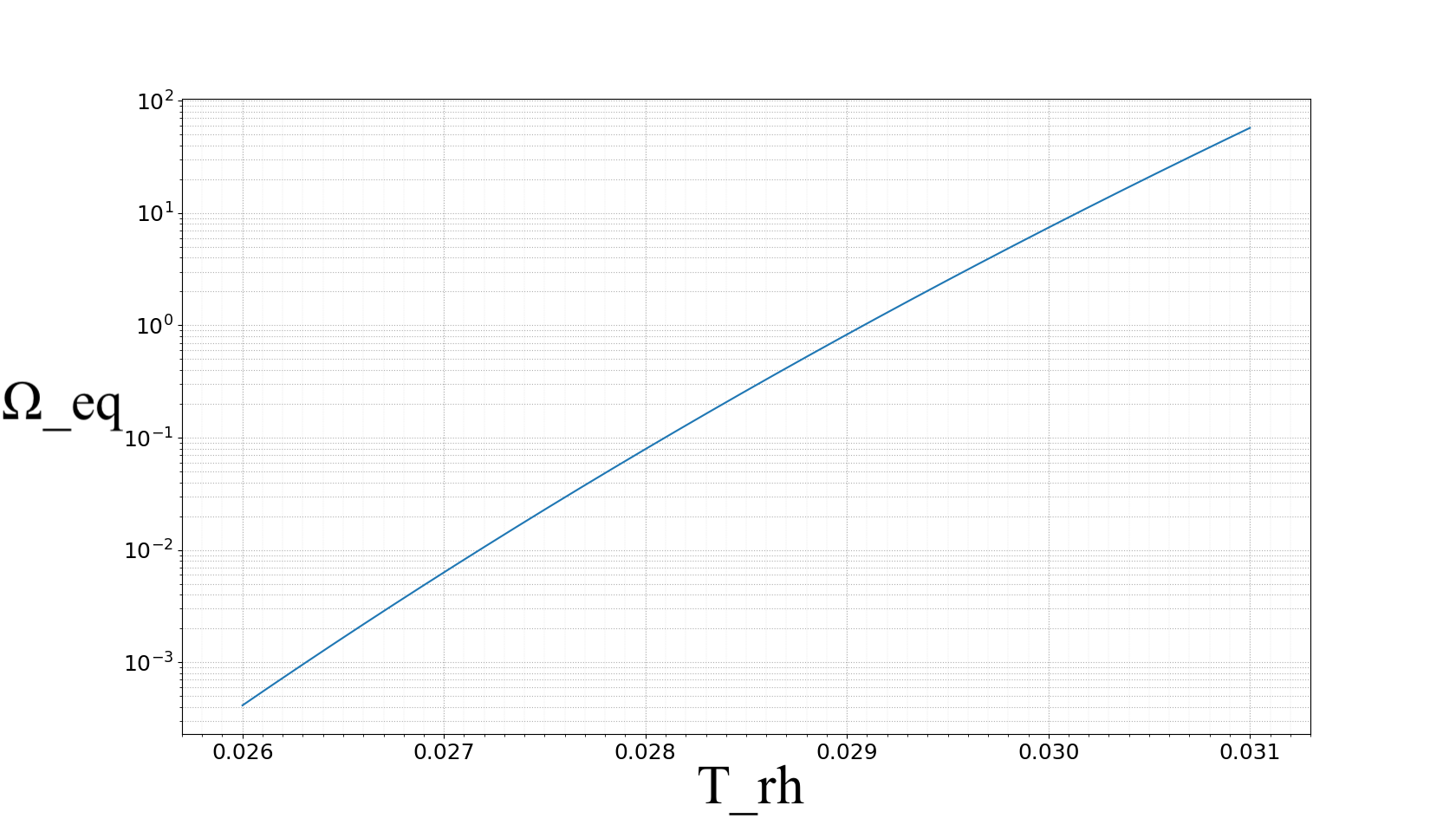}  
 \caption{Zoom on the fraction of relics at the equilibrium time as a function of the reheating temperature (in Planck units), around the relevant zone.} 
      \label{Fig2}
\end{figure}

Let us first assume that the cross section is of order 1 in Planck units $(\sigma \sim A_{Pl})$ above the threshold and that the mass of the relics is also of order 1 in Planck units $(m_{rel}\sim m_{Pl})$. In Fig \ref{Fig1}, the relative abundance of relics at the equilibrium time is plotted at the function of the reheating temperature. Figure \ref{Fig2} is a zoom on the relevant region. For a reheating temperature slightly above $10^{-2}$, one is led to a density of relics that can account for dark matter. 

\begin{figure} 
\centering
    \includegraphics[width=1.1\linewidth]{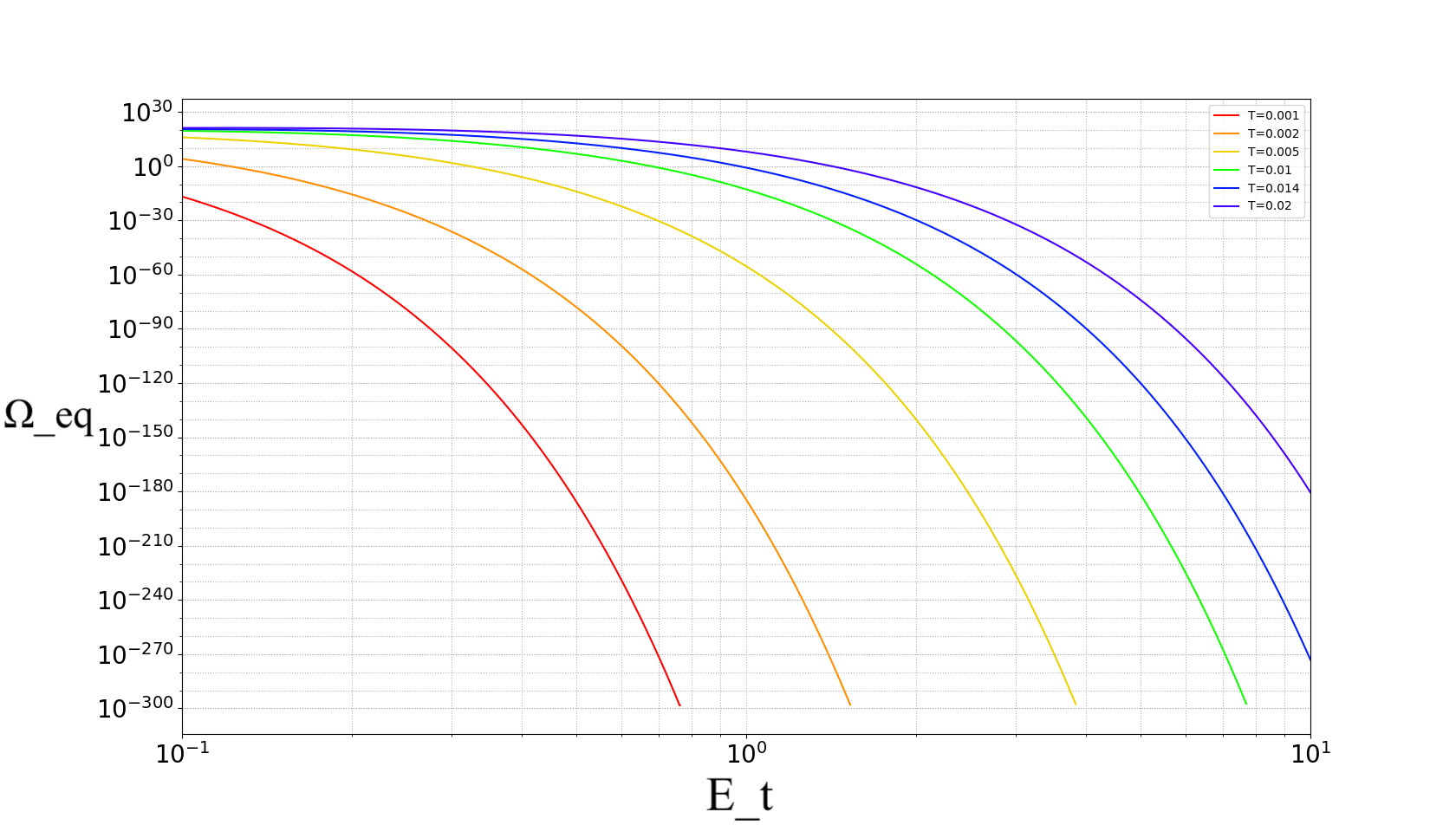}  
 \caption{Fraction of relics at the equilibrium time as a function of the energy threshold for different reheating temperatures (in Planck units), from $10^{-3}$ to $2\times 10^2$ from bottom to top.} 
      \label{Fig3}
\end{figure}

Although the influence is negligible, from now on we use the cross section $\sigma(s)=F(s)\pi R_S^2(s)$, where $R_S=2s$. We set $F=1$ above the threshold, but the dependency being linear it is easy to extrapolate to any reasonable value. In Fig. \ref{Fig3}, we show the influence of the threshold energy. The influence of the threshold energy is -- as expected -- very large. Interestingly, if nonperturbative effects were to lower the threshold by one order of magnitude with respect to the expected value, a reheating temperature around the GUT scale would be enough to produce the required density of remnants. \\

It is worth noticing that in the case with extra dimensions \cite{Nakama:2018lwy}, the ``allowed" parameter space is defined by ensuring that the Hubble rate during inflation $H_i$, together with the maximal temperature are smaller than the D-dimensional Planck scale. Meanwhile $H_i$ must remain much larger than the nucleosynthesis temperature. The formed relics account for dark matter basically between a Hubble rate of $10^{-16}$ (in usual four-dimensional Planck units) for a D-dimensional Planck scale of $10^{-7}$ to a Hubble rate of $10^{-3}$ for a D-dimensional Planck scale equal to $M_{Pl}$.

\section{The fine-tuning issue}

The model presented here seems to require a high level of fine-tuning. In particular, as the dependency upon the reheating temperature is exponential, varying slightly its value leads to a large variation in the density of relics. 
One can easily check that
$$
\frac{dT_{RH}}{T_{RH}} = \frac{d\Omega^{rel}_{eq}}{\Omega^{rel}_{eq}}\left(\frac{2E_T}{T_{RH}}+3\right)^{-1} \approx \frac{T_{RH}}{2E_T}\frac{d\Omega^{rel}_{eq}}{\Omega^{rel}_{eq}} \sim 10^{-4},
$$
to remain in agreement with data.
Unquestionably, the model requires a very high level of fine-tuning. \\

The question of fine-tuning is, however, tricky. It is only well defined relatively to an {\it a priori} specific state. In the cosmological framework, the value $\Omega=1$ is clearly such a special case. As is well known, inflation fixes a vanishing (or nearly so) curvature. Basically, as $(\Omega^{-1}-1)=-\frac{3k}{8\pi \rho a^2}$ with $\rho$ remaining constant and the scale factor increasing by at least 60 e-folds, $\Omega$ is fixed (close) to 1 at the end of inflation. There is obviously nothing magical here as $\Omega$ involves a normalization to the critical density that, itself, depends on the Hubble parameter. Should the content of the Universe be different, we would still have $\Omega=1$, with a different expansion rate. This means that changing the parameters of this model would, in fact, not drive the Universe out of the specific situation $\Omega=1$. In this sense it does {\it not} require fine-tuning. \\

One might argue that if the reheating temperature were different, other parameters of the Universe -- {\it e.g.} the (unnormalized) matter density, the equilibrium time, etc. -- would be different. This is correct. But, in our opinion, this is not a fine-tuning issue. This is just the obvious statement that things could have been different and that changing parameters do, of course, change the final state. This is not problematic as long as the ``lost" state was not a very peculiar one.\\

To summarize, the parameters of the model need to be fine-tuned so that the relic density closes the Universe at the equilibrium time -- which is a contingent fact -- but not to ensure $\Omega=1$ which is indeed the {\it a priori} specific feature. 

\section{Detectability}

Testing this model is challenging. A Planck relic has the weight of a grain of dust and no other interaction than gravity to reveal itself to the outer world. Even though the Planck mass is very small from the gravitational viewpoint, it is very large from the particle physics viewpoint. The number density of relics is therefore extremely small, even if they are to account for all the dark matter. A density of $10^{-18}$ relics par cubic meter -- that is one relic per volume of a million times the one of planet Earth -- is enough to close the Universe. Detection seems hopeless. The cross section (or greybody factor) hopefully does {\it not} tend to 0 for the absorption of fermions in the low-energy limit \cite{MacGibbon:1990zk}. However, even avoiding this catastrophic suppression (which does exist for higher spins), the area involved is of the order of the Planck one, $10^{-66}$cm$^2$, which indeed makes  direct detection  impossible in practice. \\

We consider here another possibility associated with the coalescences of relics that have occurred during the history of the Universe. Contrarily to what is sometimes done for PBHs we shall not focus on the emission of gravitational waves whose amplitude would be negligible and frequency way too high for any detector. However, something else is also expected to happen in this model. When two remnants merge, a higher-mass black hole is formed and evaporates until it reaches again $m_{rel}\sim m_{Pl}$ assumed to be the minimal one. This should happen preferably via the emission of one (or a few) quantum close to the Planck energy. Each merging should therefore emit about a Planck-energy particle which is, in principle, detectable. This sketch should of course be refined but the hypothesis is realistic enough to investigate whether this path is potentially fruitful.\\

We estimate the merging rate following \cite{Sasaki:2016jop}, which builds on \cite{Nakamura:1997sm}. It is not hard to show that the probability of coalescence in the time interval $(t,t+dt)$ is given by
$$
dP=
    \frac{3}{58} \bigg[ -{\left( \frac{t}{T} \right)}^{3/8}
+{\left( \frac{t}{T} \right)}^{3/37} \bigg] \frac{dt}{t},
$$
where $T \equiv {\bar x}^4 \frac{3}{170} {(Gm_{rel})}^{-3}$, 
$$
{\bar x}={\left( \frac{M_{rel}}{\rho_{rel}(z_{\rm eq})} \right)}^{1/3}
=\frac{1}{(1+z_{\rm eq})} 
{\left( \frac{8\pi G}{3H_0^2} \frac{m_{rel}}{\Omega_{rel}} \right)}^{1/3}
$$
being the mean separation of relics at the equilibrium time. In those formulas, we have reinserted the constants to make the use easier. The event rate is then given by
$$
n_{merg}=
\frac{3H_0^2}{8\pi G} \frac{\Omega_{rel}}{m_{rel}} 
\frac{dP}{dt}\bigg|_{t_0}.
$$
This is of the order of $10^{-45}{\rm m}^{-3}{\rm s}^{-1}$. It is then straightforward to estimate the measured flux on a detector of surface $S_d$ and solid angle acceptance $\Omega_{acc}$, integrated up to a distance $R_{max}$:
$$
\Phi_{mes}=\int_{0}^{R_{max}}n_{merg}S_d\frac{\Omega_{acc}}{4\pi}dR.
$$
Although it is well known that TeV photons are absorbed by interactions with the infrared background and PeV photons by interactions with the CMB photons, there is no strong absorption to be expected for Planck-energy photons. The wavelength of the background photons that would lead to a center-of-mass energy close to the electron mass is way larger than any expected background. The $R_{max}$ value can therefore be assumed to be much larger than for usual high-energy cosmic-ray estimations. For detectors like Auger \cite{Covault:2019mdm}, the expected flux is too small for a detection. For Euso-like instruments \cite{Inoue:2009zz} -- looking at the atmosphere from the space station -- we are led to an order of magnitude not far from a fraction of an event per year. For speculative ideas about using giant planets as cosmic-ray detectors \cite{Rimmer:2014cia}, we reach a dozen events per year. This is obviously a hard task but, interestingly, the model is clearly not unfalsifiable. \\

Furthermore, the idea that the Hawking radiation due to the formed black holes before they become stable relics might play a role in baryogenesis was considered in \cite{Alexander:2007gj}. The possibility that they might have an effect on the primordial nucleosynthesis should also be considered. In the case considered in this article -- with a true Planck scale at the four-dimensional value --, the relics are so heavy and heir number density so small that it is easy to check that the associated signal would be entirely negligible. 

\section{Conclusion}

The idea that dark matter could be made of Planck relics is not new. Nor is the possibility that black holes could be formed by the scattering of trans-Planckian particles in the early Universe. In this article we have gathered all the ingredients and argued that the resulting model is not (that) exotic. Unquestionably, the very high reheating temperature required raises questions. We have however explained that the upper bounds usually considered can be circumvented. Still, building a consistent cosmological model with such a high scale for inflation is not trivial and should be considered as a challenge.\\

There is no obvious solution to the dark matter problem, which is one of the oldest enigmas of contemporary cosmology. The scenario suggested here is based on a minimum amount of ``new physics", if not only on known physics. It requires a quite unusual cosmological behavior but no new particle physics input is needed. From this point of view, it might be worth being considered seriously.\\

Several developments would be worth being considered:
\begin{itemize}
\item the possibility that nonthermal processes do happen during the reheating, eventually enhancing the high-energy tail of the distribution should be studied with care.
\item the model presented in this article should be investigated in the context of noninflationary bouncing cosmologies, expected to be more favorable to this scenario. 
\item the experimental possibilities that were outlined should be made more precise thanks to Monte Carlo simulations.
\item in addition to the possible detection of extremely high-energy  gamma rays, it might be interesting to consider a possible low-energy (in the 100 MeV range) signal due to the disintegration of neutral pions produced by the hadronization of quarks or gluons emitted by the merging of relics.
\end{itemize}

\section{Acknowledgments}

K.M is supported by a grant from the CFM foundation. 
\bibliography{refs}

 \end{document}